\documentclass[twocolumn,english,aps,prb,floatfix,superscriptaddress]{revtex4}
\usepackage{ae,aecompl}

\usepackage[T1]{fontenc}
\usepackage[latin9]{inputenc}
\usepackage{float}
\usepackage{booktabs}
\usepackage{amsmath}
\usepackage{graphicx}
\usepackage{amssymb}

\makeatletter

\providecommand{\tabularnewline}{\\}

\@ifundefined{textcolor}{}
{%
 \definecolor{BLACK}{gray}{0}
 \definecolor{WHITE}{gray}{1}
 \definecolor{RED}{rgb}{1,0,0}
 \definecolor{GREEN}{rgb}{0,1,0}
 \definecolor{BLUE}{rgb}{0,0,1}
 \definecolor{CYAN}{cmyk}{1,0,0,0}
 \definecolor{MAGENTA}{cmyk}{0,1,0,0}
 \definecolor{YELLOW}{cmyk}{0,0,1,0}
 }

\usepackage{hyperref}\usepackage{epsfig}
\@ifundefined{definecolor}
 {\usepackage{color}}{}
\usepackage{bm}\usepackage{subfigure}

\makeatother

\usepackage{babel}

\begin{document}

\title{Zero-temperature dynamics of solid $^{4}$He \\
 from quantum Monte Carlo simulations }

\author{Giuseppe Carleo}

\affiliation{SISSA -- Scuola Internazionale Superiore di Studi Avanzati, via Beirut
2-4, I-34014 Trieste, Italy}

\affiliation{CNR-INFM DEMOCRITOS National Simulation Center, via Beirut 2-4, I-34014
Trieste, Italy}

\author{Saverio Moroni}

\affiliation{CNR-INFM DEMOCRITOS National Simulation Center, via Beirut 2-4, I-34014
Trieste, Italy}

\affiliation{SISSA -- Scuola Internazionale Superiore di Studi Avanzati, via Beirut
2-4, I-34014 Trieste, Italy}

\author{Stefano Baroni}

\affiliation{SISSA -- Scuola Internazionale Superiore di Studi Avanzati, via Beirut
2-4, I-34014 Trieste, Italy}

\affiliation{CNR-INFM DEMOCRITOS National Simulation Center, via Beirut 2-4, I-34014
Trieste, Italy}

\date{\today}
\begin{abstract}
The lattice dynamics of \emph{hcp} crystalline $^{4}$He is studied
at zero temperature and for two different densities (near and far
from melting), using a ground-state path-integral quantum Monte Carlo
technique. The complete phonon dispersion is obtained, with particular
attention to the separation of optic and acoustic branches and to
the identification of transverse modes. Our study also sheds light
onto the residual coherence affecting quasiparticle excitations in
the intermediate momentum region, in-between the phonon and nearly-free
particle regimes. 
\end{abstract}


\maketitle
The lattice dynamics of solid $^{4}$He has long been considered a
major challenge to \emph{ab initio} calculations, due to the strong
anharmonicity of this highly quantum solid. Many calculations have
been performed within the Self Consistent Phonon (SCP) approximation,\cite{Gillis:1968fk,GlydeBook}
which however may be rather unsatisfactory, due to the magnitude of
anharmonic effects. A significant improvement has been made possible
by the application of a variational quantum Monte Carlo (QMC) approach,
based on the \emph{shadow wave function} formalism, to \textit{bcc}
$^{3}$He~\cite{GalliTransverse} and to \textit{hcp} $^{4}$He.\cite{Galli:2003lr}
The variational nature of this approach, however, makes it not fully
suitable at high energy where optical or zone-boundary longitudinal
excitations exhibit broad multiphonon features. In this spectral regime,
the calculation of the full dynamic structure factor is therefore
in order. QMC techniques based on path integrals allow quite naturally
for the calculation of imaginary-time correlation functions from which
various spectral functions, such as the dynamic structure factor,
can be obtained upon analytical continuation. This technique has been
successfully demonstrated for superfluid $^{4}$He,\cite{Boninsegni:1996lr,Baroni:1999lr}
as well as for the \textit{bcc} crystalline phases of $^{4}$He~~\cite{Sorkin:2005lr,Pelleg:2008lr}
and $^{3}$He.\cite{Sorkin:2006fk} In the latter studies the spectrum
of the transverse excitations has also been obtained, albeit in the
one-phonon approximation only.\cite{GlydeBook}

In this work we present an extensive study of the dynamical properties
of \textit{hcp} $^{4}$He at zero temperature, performed by estimating
the dynamic structure factor from ground-state path-integral simulations.
\cite{Ceperley:1995fk,Baroni:1999lr,Sarsa:2000lr} This technique
allows us to parallel to some extent the procedure followed experimentally
to map phonon dispersions from the measured neutron scattering. In
the long wave-length region---well approximated by a phonon picture
of the collective density excitations---we thus obtain longitudinal
as well as transverse modes for both acoustic and optical branches.
For higher wave-vectors we analyse the dynamic structure factor in
terms of corrections to the so-called \emph{impulse approximation},\cite{Glyde:1994fk}
finding a coherent response which is peculiar of both superfluid and
solid helium.

In Sec. \ref{sec:Latticedyn} we give an introductory account of the
phonon theory of long wave-length excitations in solids. In Sec. \ref{sec:NumMeth}
the reader is provided with an outline of the numerical methods adopted
in this work. In Sec. \ref{sec:Results} we report on the analysis
of our QMC results both in the phonon regime and in the intermediate
momentum region. Sec. \ref{sec:Conclusions} is finally devoted to
a few concluding remarks.


\section{\label{sec:Latticedyn}Lattice dynamics}

\subsection{Long wave-lengths}

The long wave-length lattice dynamics of a solid is fully characterized
by its dynamic structure factor, which is the space-time Fourier transform
of the density-density correlation function. In real time and reciprocal
space, the autocorrelation function of the density operator reads:
\begin{equation}
S(\mathbf{Q},t)=\frac{1}{N}\left\langle \sum_{kl}e^{-i\mathbf{Q}\cdot\mathbf{r}_{k}(t)}e^{i\mathbf{Q}\cdot\mathbf{r}_{l}(0)}\right\rangle ,\end{equation}
where the brackets indicate equilibrium (ground-state or thermal)
expectation values. In a weakly anharmonic system, it is convenient
to expand $S(\mathbf{Q},t)$ into a sum of terms involving one-phonon
processes, two-phonon scattering, interference processes and so on:\cite{GlydeBook}
\begin{equation}
S(\mathbf{Q},t)=S_{1}(\mathbf{Q},t)+S_{2}(\mathbf{Q},t)+S_{1,2}(\mathbf{Q},t)+\cdots.\label{expansion}\end{equation}
The physical meaning of such an expansion is best appreciated by introducing
the atomic displacements from the equilibrium lattice sites, $\{\mathbf{R}_{l}\}$:
$\mathbf{u}_{l}(t)=\mathbf{r}_{l}(t)-\mathbf{R}_{l}$. In terms of
the $\mathbf{u}$'s and the $\mathbf{R}$'s, the one-phonon contribution
to the dynamic structure factor of a simple Bravais lattice reads:\cite{GlydeBook}
\begin{multline}
S_{1}(\mathbf{Q},t)=e^{-2W}{\sum_{l}e^{-i\mathbf{Q}\cdot\left(\mathbf{R}_{l}-\mathbf{R}_{0}\right)}}\times\\
\times\left\langle \mathbf{Q}\cdot\mathbf{u}_{l}(t)\mathbf{Q}\cdot\mathbf{u}_{0}(0)\right\rangle ,\end{multline}
where $e^{-2W}$ is the Debye-Waller factor. For a harmonic crystal---to
which only, strictly speaking, the phonon language applies---we consider
the vibrational frequency $\omega_{j}(\mathbf{q})$ and polarization
vector $\boldsymbol{\epsilon}(\mathbf{q}|j)$ of the $j$-th phonon
branch at wave vector $\mathbf{q}$ in the first Brillouin Zone (FBZ).
In terms of these quantities, the one-phonon contribution reads:\cite{Brockhouse:1958lr}
\begin{equation}
S_{1H}(\mathbf{Q},t)=\sum_{j}g^{2}(\mathbf{Q}|j)e^{-i\omega_{j}(\mathbf{q})t},\label{eq:multiS}\end{equation}
 where $\mathbf{Q}=\mathbf{q}+\mathbf{G}$, $\mathbf{G}$ being a
reciprocal-lattice vector, and $g^{2}(\mathbf{Q}|j)\propto\left|\mathbf{Q}\cdot\boldsymbol{\epsilon}(\mathbf{Q}|j)\right|^{2}$
is the so-called inelastic structure factor that filters out transverse
vibrations. In the case of a non-Bravais lattice, such as the \textit{hcp}
phase of Helium, the form of the inelastic structure factor is slightly
more complicated:\cite{Brockhouse:1958lr} \begin{equation}
g^{2}(\mathbf{Q}|j)=e^{-2W}\frac{\hbar}{2m\omega_{j}(\mathbf{q})}\left|\sum_{k}\mathbf{Q}\cdot\boldsymbol{\epsilon}_{k}(\mathbf{q}|j)e^{i\mathbf{Q}\cdot\mathbf{d}_{k}}\right|^{2},\label{eq:gquadro}\end{equation}
 where the $\mathbf{d}$'s are the positions of the atomic basis.

In a perfectly harmonic solid the Fourier transform of Eq. \eqref{eq:multiS},
$S_{1H}(\mathbf{Q},\omega)$ is merely a sum of Dirac delta functions
centered at the phonon frequencies. 
In a real solid, things are more complicated: anharmonic interactions
broaden the one-phonon peaks and give rise to non-vanishing multi-phonon
and interference contributions to the dynamic structure factor (Eq.
\ref{eq:multiS}). When anharmonic effects are not too large, 
one-phonon excitations can still be long-lived---thus providing a
reasonable description of the dynamics---and 
it is thus well justified to identify the positions of the finite-width
peaks of $S_{1}(\mathbf{Q},\omega)$ with phonon frequencies. From
an experimental point of view, phonon frequencies are generally extracted
from the peaks of the full dynamic structure factor $S(\mathbf{Q},\omega)$.
The cross section of inelastic neutron or X-ray scattering is in fact
proportional to $S(\mathbf{Q},\omega)$ \cite{PhysRev.95.249} and
no direct access is possible to its one-phonon component. The latter
dominates the cross section only at small transferred momentum, whereas
multi-phonon contributions cannot in general be neglected when pursuing
a comparison between calculated and measured phonon dispersions.

\subsection{Shorter wave-lengths}

The very concept of \emph{phonon}, which lies at the basis of the
theory of lattice dynamics sketched above, is most appropriate to
describe the low-lying portion of the spectrum of solid $^{4}$He,
probed by inelastic neutron or X-ray scattering at long wave-lengths.
In the opposite limit of short wave-lengths, the scattering process
can be pictured as the creation of particle-hole pairs, resulting
from the high momentum transferred to the crystal from the incoming
particle beam.\cite{GlydeBook} The kinematics of the struck particles
in the higher-energy states is clearly affected by the distribution
of allowed atomic momenta, $n(\mathbf{Q})$, and neutron spectroscopy
at large momentum transfer has in fact proven useful to probe off-diagonal
long-range order, both in superfluid \cite{Sears:1982kl} and, more
recently, in solid Helium.\cite{Diallo:2007qy}

At intermediate wave-lengths both the phonon and a purely impulsive,
particle-hole, picture of density excitations break down. In spite
of the attention paid by both experimentalists and theorists to this
peculiar intermediate regime, both in the superfluid \cite{Martel:1976db,Tanatar:1987rw}
and in the solid\cite{Glyde:1985lr,Diallo:2004lr} phases, it turns
out that the neglect of interaction-induced coherence effects make
previous theoretical studies not totally satisfactory.\cite{Glyde:1985lr} 

At small wave-length, the solid behaves like a collection of almost
non-interacting atoms and the intermediate scattering function can
be approximated by its incoherent part, \cite{Glyde:1994fk} i.e.

\begin{equation}
S_{\text{inc}}(\mathbf{Q},t)=\frac{1}{N}\left\langle \sum_{l}e^{-i\mathbf{Q}\cdot\mathbf{r}_{l}(t)}e^{i\mathbf{Q}\cdot r_{l}(0)}\right\rangle ,\label{eq:Sinco}\end{equation}
which amounts to neglecting the interference terms involving different
atoms. For a crystal, the incoherent part can be expressed in terms
of the recoil frequency $\omega_{R}=\frac{\hbar}{2m}Q^{2}$ and of
the phonon density of states $g(\omega)$, leading to

\begin{equation}
S_{\text{inc}}(\mathbf{Q},t)=\exp\left[\omega_{R}\int_{0}^{\infty}d\omega g(\omega)\frac{1}{\omega}(e^{-i\omega t}-1)\right].\end{equation}
Such an expression has been used by Glyde \cite{Glyde:1985lr} to
compute the incoherent response of \textit{bcc} $^{4}$He. By its
very nature, the incoherent approximation is only reliable at very
high wave-vector,\cite{Glyde:1985lr} roughly larger than $20\,\text{\AA}^{-1}$.
In order to account for the leading coherence effects on the short
wave-length dynamics of an extended system, it is convenient to consider
a cumulant expansion of the intermediate scattering function,\cite{Glyde:1994fk}

\begin{equation}
S(\mathbf{Q},t)=S(\mathbf{Q})e^{-i\omega_{R}t}\exp\left[\sum_{n=1}^{\infty}\frac{\mu_{n}}{n!}(-it)^{n}\right],\label{eq:cumul}\end{equation}
where $S(\mathbf{Q})$ is the static structure factor and $\mu_{n}$
are the cumulants of the distribution $S(\mathbf{Q},\omega-\omega_{R})$.
Retaining the leading contribution to such an expansion yields the
so-called \emph{Impulse Approximation}, according to which the Fourier
transform of the dynamic structure factor consists of a main Gaussian
component centered at the recoil frequency $\omega_{R}$, plus additive
corrections: \cite{Azuah:1997sf}

\begin{multline}
S(\mathbf{Q},\omega)=\tilde{S}_{IA}(\mathbf{Q},\omega)+\tilde{S}_{1}(\mathbf{Q},\omega)+\\
+\tilde{S}_{2}(\mathbf{Q},\omega)+\tilde{S}_{3}(\mathbf{Q},\omega)+\cdots,\label{eq:cumulomega}\end{multline}
 where the first terms of the expansion read

\begin{eqnarray}
\tilde{S}_{IA}(\mathbf{Q},\omega) & = & \frac{S(\mathbf{Q})}{\sqrt{2\pi\mu_{2}}}e^{-\frac{\omega_{d}^{2}}{2}}\nonumber \\
\tilde{S}_{1}(\mathbf{Q},\omega)\,\, & = & -\frac{\mu_{3}}{2\mu_{2}^{2}}\left(\omega-\omega'_{R}\right)\left[1-\frac{\omega_{d}^{2}}{3}\right]\tilde{S}_{IA}(\mathbf{Q},\omega)\nonumber \\
\tilde{S}_{2}(\mathbf{Q},\omega)\,\, & = & \frac{\mu_{4}}{8\mu_{2}^{2}}\left[1-2\omega_{d}^{2}+\frac{\omega_{d}^{4}}{3}\right]\tilde{S}_{IA}(\mathbf{Q},\omega)\nonumber \\
\tilde{S}_{3}(\mathbf{Q},\omega)\,\, & = & \frac{\mu_{5}}{8\mu_{2}^{3}}\left(\omega-\omega'_{R}\right)\times\nonumber \\
 &  & \times\left[1-\frac{2}{3}\omega_{d}^{2}+\frac{\omega_{d}^{4}}{15}\right]\tilde{S}_{IA}(\mathbf{Q},\omega)\label{eq:ExplCumul}\end{eqnarray}
 with $\omega'_{R}=\omega_{R}/S(\mathbf{Q})$ and $\omega_{d}^{2}=\left(\omega-\omega'_{R}\right)^{2}/\mu_{2}$
.



\section{Numerical methods \label{sec:NumMeth}}

The dynamical properties of bosonic systems are conveniently simulated
in imaginary time, $\tau=it$, using path integral QMC methods, at
both finite \cite{Boninsegni:1996lr} and zero \cite{Baroni:1999lr}
temperature. Specializing to the zero temperature case, a discretized
path integral expression for the imaginary-time propagator can be
used to project out the exact ground state $\Psi_{0}$ from a positive
trial wave function $\Phi_{0}$, according to

\begin{equation}
\Psi_{0}=\lim_{\beta\to\infty}\exp(-\beta\widehat{H})\Phi_{0},\end{equation}
 thus mapping the imaginary-time evolution, from which ground-state
expectation values can be obtained, onto a classical system whose
fundamental variables are open quantum paths (or \emph{reptiles} in
the parlance of Refs. \onlinecite{Baroni:1999lr} and \onlinecite{ProcRept}).
Full details of the formalism can be found in Ref.~\onlinecite{ProcRept}
and need not be repeated here; we only stress that unbiased ground-state
expectation values are obtained when the projection time $\beta$
is large enough and the step $\epsilon$ of the time discretization
is small enough. 
We simulate samples of $^{4}$He atoms interacting through the Aziz
\cite{Aziz79} pair potential, placed in a cuboid cell accommodating
an \textit{hcp} lattice. The number $N$ of particles is either $180$
or $360$, the latter corresponding to a cell with double extension
in the $\Gamma A$ direction. Altough all the reported results refer
to the larger system, we have found a full agreement between the relevant
observables computed on the common set of wave vectors shared by the
smaller and the larger simulation cells. For systems of this size,
we find it more efficient to use the bisection algorithm \cite{Ceperley:1995fk,Sarsa:2000lr}
rather than the reptation algorithm \cite{Baroni:1999lr} for sampling
the path space. We adopt the so-called \textit{primitive approximation}
for the imaginary-time propagator,\cite{Ceperley:1995fk} which requires
a small time step $\epsilon=10^{-3}$ inverse $K$ for accurate results,
and we set the projection time to $\beta=0.3$. The correlation functions
of the density operators are calculated for values of $\tau$ up to
$\tau_{{\rm max}}=0.25$, implying that our paths span an imaginary
time of $2\beta+\tau_{{\rm max}}=0.85$. 

The trial function is of the standard McMillan-Nosanow form,\begin{eqnarray}
\Phi_{0}(R) & = & \exp\left[-\sum_{k<l}ar_{kl}^{-b}-\sum_{l}c({\bf r}_{l}-{\bf s}_{l})^{2}\right]\label{eq:trialwf}\end{eqnarray}
 where $\{{\bf r}_{1},\ldots,{\bf r}_{N}\}\equiv R$ are the coordinates
of the $N$ atoms, and ${\bf s}_{l}$ is the $l$-th site of the \textit{hcp}
lattice. The three parameters $a$, $b$, and $c$ are optimized by
minimizing the variational energy and their numerical values for the
densities considered in this paper are shown in Table \ref{tab:Optimal-values-of}.
The Gaussian localization terms in the trial function explicitly break
Bose symmetry at the variational level. Nonetheless, we believe that
the lack of permutation simmetry cannot affect the determination of
the density excitation energies due to the small frequency of particle
exchanges in the crystal. This point can be further elucidated noticing
that the presence of a vacancy in the simulation box greatly enhances
the number of exchanges.\cite{Clark:2008bf} Therefore, if indistinguishability
were important, one would expect a change in the frequency upon doping
with vacancies. However, no such effect was found in in a variational
calculation\cite{Galli:2003lr} where Bose symmetry was taken into
account.%
\begin{table}[H]
\begin{centering}
\begin{tabular}{cccc}
\toprule 
\addlinespace
$\rho\,\left[{\text{Å}}^{-3}\right]$ & $a\left[{\text{Å}}^{b}\right]$ & $b$ & $c\left[{\text{Å}}^{-2}\right]$\tabularnewline\addlinespace
\midrule
\midrule 
$0.028$ & $156.0$ & $5.74$ & $0.565$\tabularnewline
\midrule 
$0.037$ & $139.4$ & $5.75$ & $0.993$\tabularnewline
\bottomrule
\end{tabular}
\par\end{centering}

\caption{Optimal values of the variational parameters appearing in the McMillan-Nosanow
wave-function. \label{tab:Optimal-values-of}}

\end{table}

\begin{figure}[h]
\includegraphics[width=1\columnwidth]{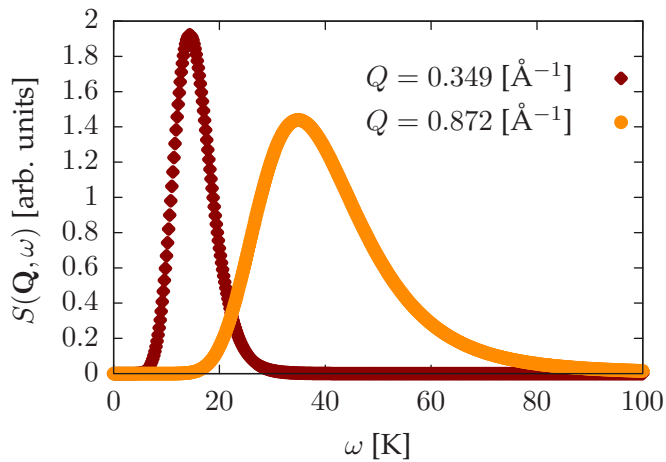} 

\caption{\label{fig:maxentcomp}(Color online) Example of two reconstructed
spectra at different wave-vectors along the $\Gamma A$ direction
at the melting density $\rho_{1}$. }

\end{figure}

In order to obtain $S(\mathbf{Q},\omega)$ from imaginary-time correlations,
an inverse Laplace transform must be performed, for which we use the
 \emph{maximum entropy} method.\cite{Gubernatis:1991fk} Although
the reconstructed spectra are typically much too broad, this procedure
gives good results, at least for the position of the peaks, when a
single sharp feature exhausts most of the spectral weight. As a typical
example, we show in Figure \ref{fig:maxentcomp} two spectra at different
wave-vectors for the longitudinal acoustic branch, calculated at the
melting density $\rho_{1}=0.028\,[{\text{\AA}}^{-3}]$. Despite the
fact that much of the width of the peaks is an artifact of the numerical
inversion of the Laplace transform, it is nonetheless plausible that
the broadening of the spectra shown in Figure \ref{fig:maxentcomp}
reflects stronger multiphonon effects at higher wave-vectors. In the
following we refer to the position of the peaks of the reconstructed
spectrum as to \emph{phonon energies}. The reported error bar is the
statistical uncertainty of the peak position, as estimated with the
jackknife resampling method.\cite{resamplmeth}%

\section{Results \label{sec:Results}}

In this Section we present the results of our QMC simulations for
both the phonon dispersion energies and the higher wave-vectors response
of the solid.

\subsection{Long wave-length excitations}

\begin{figure*}
\centering \includegraphics[width=2\columnwidth]{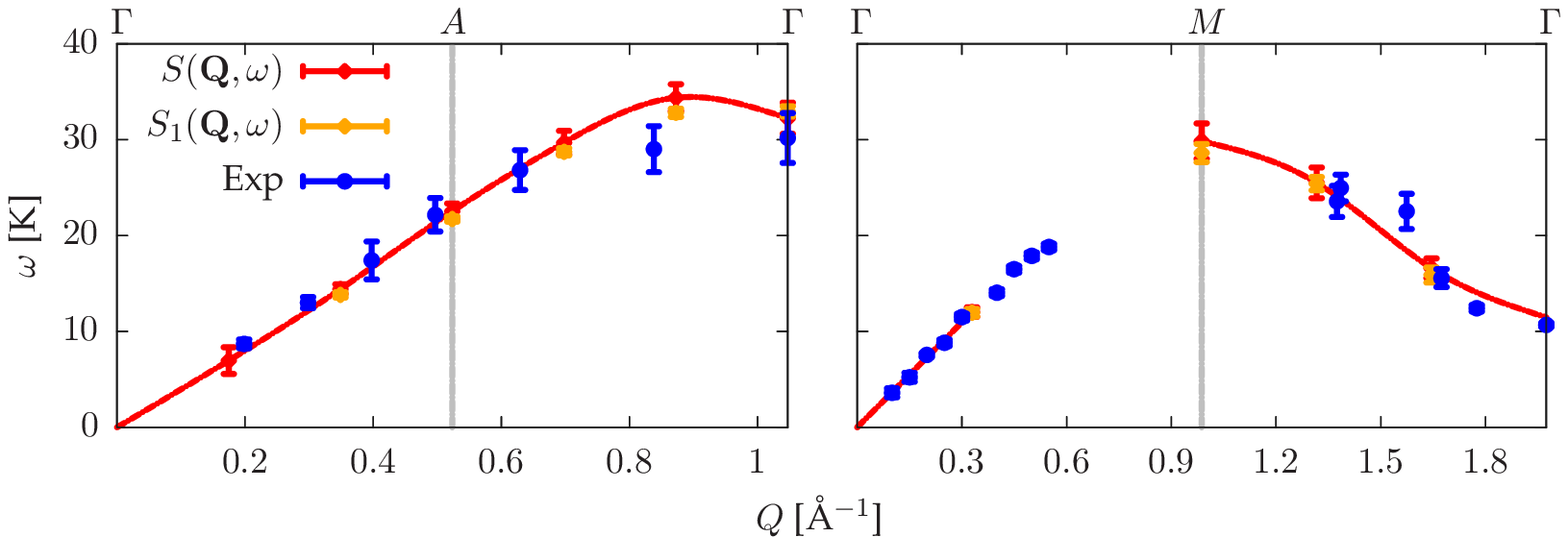}

\caption{\label{fig:rho1L} (Color online) Longitudinal phonon energies at
the melting density $\rho_{1}$ extracted from $S(\mathbf{Q},\omega)$
and $S_{1}(\mathbf{Q},\omega)$, $\Gamma A$ and $\Gamma M$ directions
(resp. left and right panel). The limit of the FBZ is indicated by
a vertical grey line. Experimental data from \onlinecite{Minkiewicz:1968lr}
and \onlinecite{Minkiewicz:1973fk}. Frequencies calculated at discrete
wave-vectors are interpolated by cubic splines as a guide to the eye.}

\end{figure*}

\begin{figure*}
\centering

\includegraphics[width=2\columnwidth]{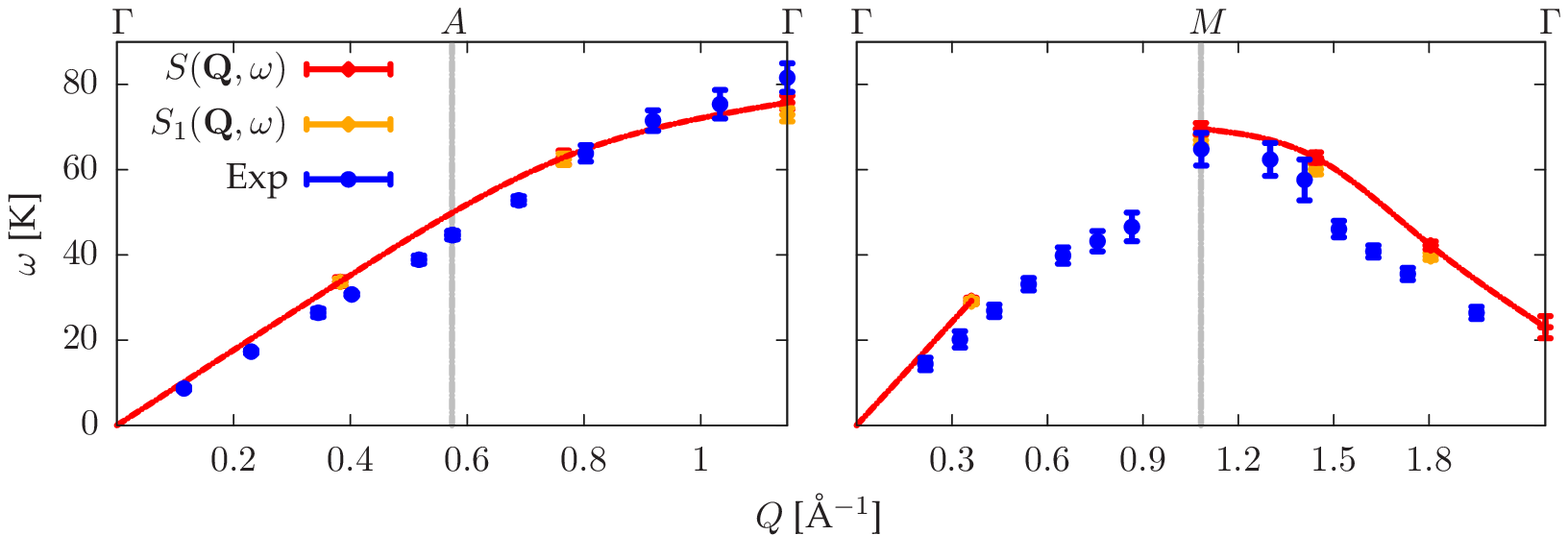} 

\caption{\label{fig:rho2L} (Color online) Longitudinal phonon energies at
the density $\rho_{2}=0.037\,[{\text{Å}}^{-3}]$ extracted from $S(\mathbf{Q},\omega)$
and $S_{1}(\mathbf{Q},\omega)$, $\Gamma A$ and $\Gamma M$ directions
(resp. left and right panel). The limit of the FBZ is indicated by
a vertical grey line. Experimental data from \onlinecite{Reese:1971qy}.
Frequencies calculated at discrete wave-vectors are interpolated by
cubic splines as a guide to the eye.}

\end{figure*}

\subsubsection*{Longitudinal modes}

The excitation energies of longitudinal vibrations can be straightforwardly
obtained from the dynamic structure factor $S(\mathbf{Q},\omega)$,
when available. In a lattice with a basis, such as \textit{hcp} $^{4}\mathrm{He}$,
multiple branches (acoustic and optic) exist at each point of the
FBZ. Although for a generic wave-vector all the branches contribute
to $S(\mathbf{Q},\omega)$, it often happens that---because of the
explicit dependence of the inelastic structure factor \ref{eq:gquadro}
on both the wave-vector and the branch index---along high-symmetry
directions different branches dominate (and in practice are only visible)
at different values of the wave-vector. As a consequence, there are
regions of the reciprocal space in which the acoustic modes dominate
while the optical modes are suppressed and vice-versa. To figure out
the relative weights of the branches, the inelastic structure factor
can be calculated in a number of (approximate) ways, as in Ref.~\onlinecite{PhysRev.128.562}
for beryllium, another \textit{hcp} solid. By virtue of the strongly
geometrical nature of $g^{2}(\mathbf{Q}|j)$, it is sufficient to
look at one of these approximate calculations performed for the \textit{hcp}
geometry to realize that, with few exceptions, the relative weights
do generally suppress one mode and privilege the other. Our results
substantially confirm this picture, the calculated spectral functions
being generally dominated by a single peak. Reconstructing a complete
picture of the phonon dispersions thus require sampling the dynamic
structure factor outside the FBZ.

We have calculated the phonon energies at the melting density $\rho_{1}=0.028\,[{\text{\AA}}^{-3}]$---in
a regime of strong quantum fluctuations signaled by a considerable
Lindemann's ratio---and at the density $\rho_{2}=0.037\,[{\text{\AA}}^{-3}]$,
where the quantum fluctuations are less pronounced. Results are shown
in Figures \ref{fig:rho1L} and \ref{fig:rho2L} using an extended-zone
scheme reminiscent of the way the optic and acoustic modes are measured
in the lab. The phonon energies extracted by $S(\mathbf{Q},\omega)$
are compared to experimental data. The phonon energies resulting from
an analysis of the one-phonon contribution to the dynamic structure
factor are also shown for comparison. 

The main findings that emerge from the calculations of the longitudinal
modes are the following: 
\begin{enumerate}
\item The overall agreement between the calculated and measured peak energies
is good. The estimated errors come from the intrinsic width of the
peaks of the dynamic structure factor, which is larger for optic than
for acoustic phonon. This feature is present in both the theoretical
and experimental spectra, although in the former the width is enhanced
by the numerical difficulties in performing inverse Laplace transforms. 
\item The discrepancy between the frequencies estimated in the one-phonon
approximation and from the full dynamic structure factor is generally
small. Multiphonon processes have clearly the effect of broadening
the spectrum---particularly for large wave vectors---but they hardly
affect the peaks' positions. 
 
\item In the $\Gamma M$ direction at the melting density $\rho_{1}$, we
obtain a substantial improvement over SCP results.\cite{Gillis:1968fk}
We are thus able to separate the optic from the acoustic branches,
which appear as distinct peaks in the dynamic structure factor. We
also obtain a significant improvement over previous variational QMC
results;\cite{Galli:2003lr} besides, in Ref.~\onlinecite{Galli:2003lr}
optical branches are calculated only in the $\Gamma A$ direction. 
\item For the higher density $\rho_{2}$ we observe a stronger discrepancy
between theoretical and experimental data, particularly in the $\Gamma M$
direction, possibly due to the pair potential adopted here.\cite{Moroni:2000fp} 
\end{enumerate}

\subsubsection*{Transverse modes}

The direct evaluation of transverse phonon modes from the dynamic
structure factor is hindered by the $\mathbf{Q}\cdot\boldsymbol{\epsilon}(\mathbf{q}|j)$
term appearing in its expression (Eqs. \ref{eq:multiS} and \ref{eq:gquadro})
that selects longitudinal modes. At least two strategies can be deployed
to circumvent this problem. The most immediate solution consists in
considering the peaks in the Fourier transform of the transverse counterpart
of the one-phonon contribution to the dynamic structure factor:\cite{Sorkin:2005lr}
\begin{equation}
S_{1\perp}(\mathbf{Q},t)\propto\left\langle \sum_{l,m}u_{\perp,l}(t)u_{\perp,m}(0)e^{i\mathbf{Q}\cdot\left(\mathbf{R}_{l}-\mathbf{R}_{m}\right)}\right\rangle ,\end{equation}
where $u_{\perp,}$ is the transverse component of the atomic displacement
from equilibrium. Although legitimate in principle, this approach
is limited to the weak anharmonic regime and only gives access to
the positions of the peaks, not to their intensities. A better approach,
which in principle also gives access to peaks intensities, is to mimic
closely the experimental practice and calculate the dynamic structure
factor at wave-vectors $\mathbf{Q}=\mathbf{G}+\mathbf{q}$ such that
$\mathbf{q}$ is arbitrarily \textit{quasi-perpendicular} to $\mathbf{Q}$,
so that a lattice vibration polarized parallel to \textbf{$\mathbf{Q}$}
is actually quasi-transverse:\cite{GalliTransverse} this is always
possible, just choosing a large enough $\mathbf{G}$, \emph{i.e.}
looking at wave-vectors in the second, third, or successive Brillouin
zones. In such a geometry transverse phonon energies at high-symmetry
wave-vectors in the FBZ can be estimated from the peaks in the dynamic
structure factor. 
 In order to achieve a close comparison between our results and experimental
data, our simulations at the melting density have been performed along
the same \textit{quasi-transverse} wave-vector directions as used
in Ref.~\onlinecite{Minkiewicz:1968lr}. 

\begin{figure*}
\centering \includegraphics[width=2\columnwidth]{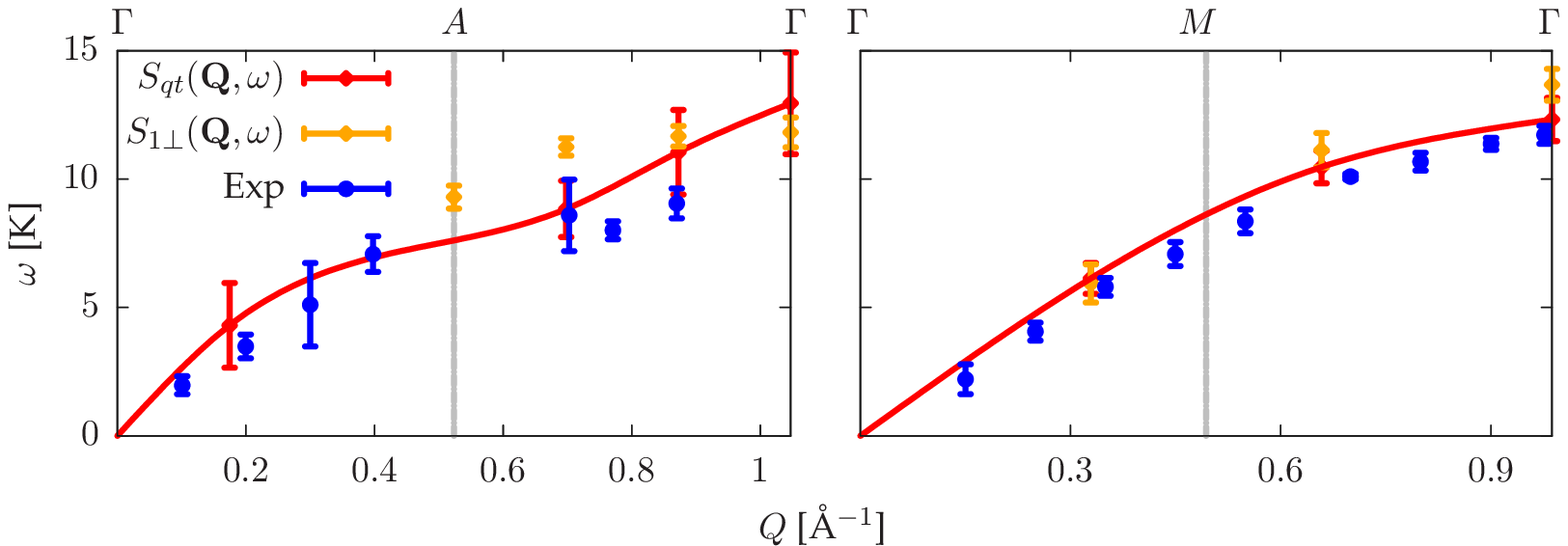}

\caption{\label{fig:rho1T} (Color online) Transverse phonon energies at the
melting density $\rho_{1}$ extracted from the \textit{quasi-transverse}
geometry of $S(\mathbf{Q},\omega)$ and from the transverse components
of $S_{1}(\mathbf{Q},\omega)$, $\Gamma A$ direction and $T_{\parallel}$
branch of the $\Gamma M$ direction (resp. left and right panel).
The limit of the FBZ is indicated by a vertical grey line. Experimental
data from \onlinecite{Minkiewicz:1968lr} and \onlinecite{Minkiewicz:1973fk}.
Frequencies calculated at discrete wave-vectors are interpolated by
cubic splines as a guide to the eye.}

\end{figure*}

\begin{figure*}
\includegraphics[width=2\columnwidth]{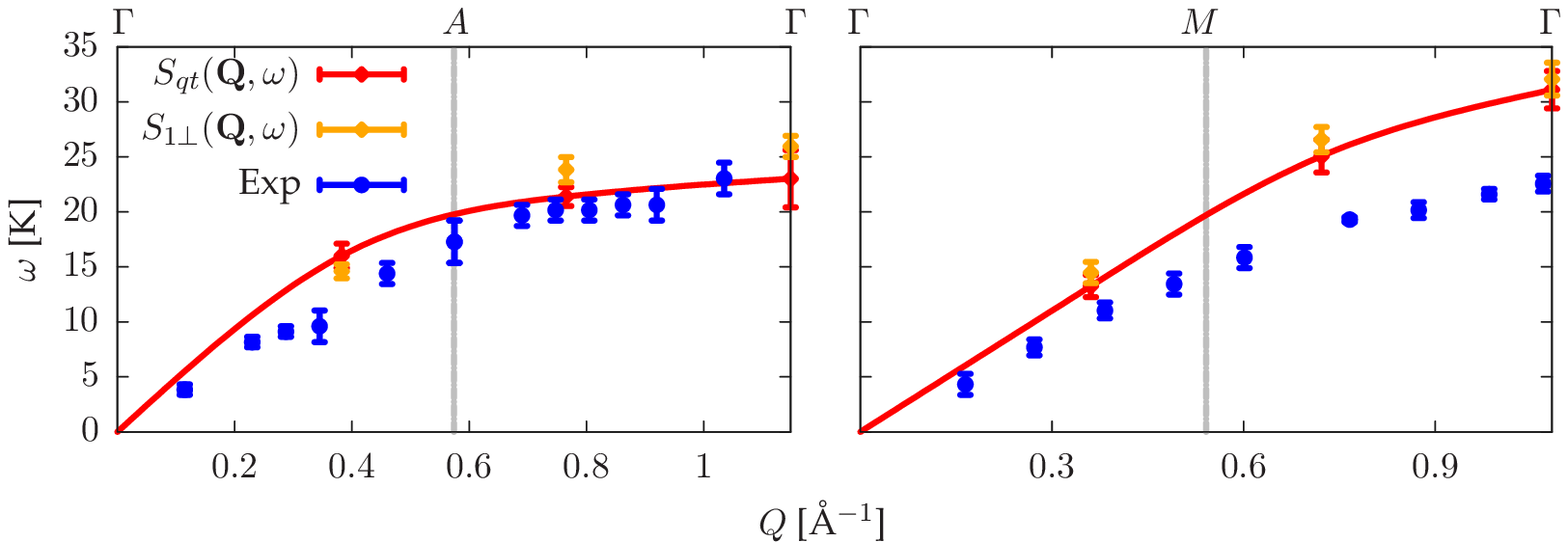}

\caption{\label{fig:rho2T} (Color online) Transverse phonon energies at the
density $\rho_{2}=0.037\,[{\text{Å}}^{-3}]$ extracted from the \textit{quasi-transverse}
geometry of $S(\mathbf{Q},\omega)$ and from the transverse components
of $S_{1}(\mathbf{Q},\omega)$, $\Gamma A$ direction and $T_{\parallel}$
branch of the $\Gamma M$ direction (resp. left and right panel).
The limit of the FBZ is indicated by a vertical grey line. Experimental
data from \onlinecite{Reese:1971qy}. Frequencies calculated at discrete
wave-vectors are interpolated by cubic splines as a guide to the eye.}

\end{figure*}

The transverse phonon energies thus obtained are presented in Figures
\ref{fig:rho1T} and \ref{fig:rho2T}, in an extended zone scheme.

The main findings that emerge from the calculations of the transverse
modes closely parallel the results obtained in the longitudinal case: 
\begin{enumerate}
\item The overall agreement with the experimental data is good. 
\item The discrepancy between the energy obtained from the full dynamic
form factor and from its one-phonon component is small. Energies from
the full form factor tend to be more noisy than in the longitudinal
case, possibly due to larger multi-phonon effects related to the large
wave-vector involved in the quasi-transverse geometry. 
\item We find a substantial improvement over the SCP results, particularly
in the $\Gamma M$ direction, whilst there are no QMC results to compare
with. 
\item We find a systematic degradation of the agreement with experimental
results for increasing density, especially for the direction $\Gamma M$
of Figure \ref{fig:rho2T}.
\end{enumerate}

\subsection{\label{sec:Hwvd}Intermediate wave-length excitations}

Recent claims that solid $^{4}$He may display a \emph{supersolid}
behavior closely related to superfluidity in the liquid phase have
prompted a revived interest in the experimental investigation of density
excitations at intermediate wave-length, from which valuable information
on the atomic momentum distribution can be extracted.\cite{Diallo:2007qy}
This experimental effort relies on the accurate determination of corrections
to the Impulse Approximation (Eq. \ref{eq:cumulomega}), a task which
is facilitated in the large-momentum regime.\cite{Glyde:1994fk} 

Apart from the issues of off-diagonal long-range order and Bose-Einstein
condensation in solid $^{4}$He, the role of atomic interference---as
emerging from the additive corrections to the bare Impulse Approximation---has
not yet been the subject of detailed theoretical investigations. To
our knowledge, the best quantitative account of the response of solid
helium is limited to the regions of small and very large wave-vectors.\cite{Glyde:1985lr}
Nonetheless in the \textit{intermediate} region where the long-wavelength
spectra of both the liquid and the solid merge into the large-momentum
regime of nearly free-particle recoil, no \emph{ab initio} results
have been reported so far. %
\begin{figure}[t]
\includegraphics[width=1\columnwidth]{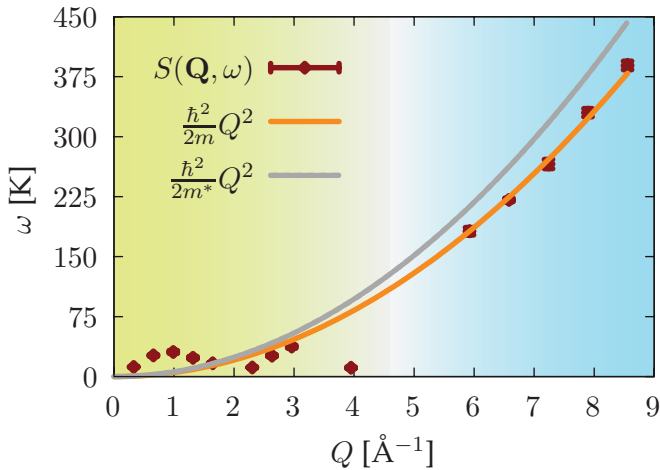}

\caption{\label{fig:yLh028} (Color online) Longitudinal excitations along
the $\Gamma M$ direction at the density $\rho_{1}$. The solid line
is the fitted free-particle dispersion of a quasi-particle with effective
mass $m*$, while the grey line is the free-particle dispersion of
atomic Helium. The two different background colors ideally separate
the phonon region from the intermediate region of wave-vectors. }

\end{figure}
\begin{figure}[H]
\includegraphics[width=1\columnwidth]{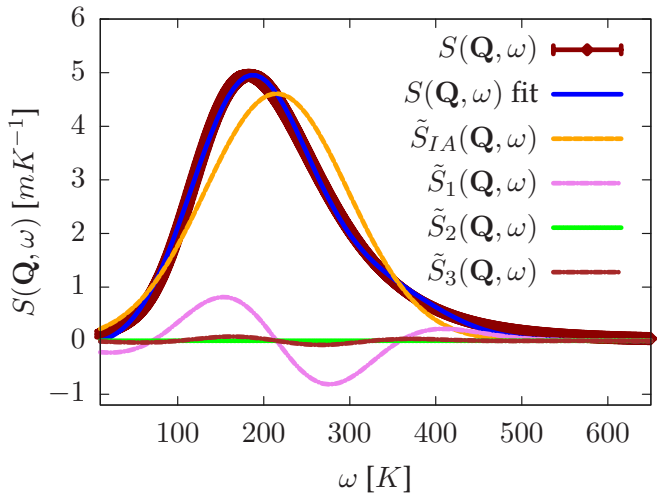} 

\caption{\label{fig:yLhfit1028} (Color online) Analysis of the dynamic structure
factor into the Impulse Approximation and its additive corrections
at the wave-vector $Q=5.92\,\text{\AA}^{-1}$ along the $\Gamma M$
direction. }

\end{figure}
Our QMC methodology, instead, allows us to provide an accurate description
of this regime as well, showing evidence of a phonon-like residual
coherence, not dissimilar to what is found in the superfluid phase.\cite{Martel:1976db}
We have concentrated our attention to wave-vectors roughly ranging
from $5\,\text{\AA}^{-1}$ to $10\,\text{\AA}^{-1}$ between the phonon
and the purely\textit{ single particle} regimes. The transition between
these two regions can be clearly seen looking at the dispersion of
the peaks of the dynamic structure factor, as a function of the excitation
wave-vector. In Figure \ref{fig:yLh028} the longitudinal excitation
energies of the crystal at the melting density $\rho_{1}$ are shown.
In the phonon-like regime (on the left of the figure with a yellowish
background) we observe a periodic dispersion with soft modes corresponding
to reciprocal-lattice vectors; for larger wave-vectors (on the right
with a bluish background) we observe a free-particle parabolic dispersion,
corresponding to an effective mass which is slightly higher than the
bare Helium mass, being $M*\simeq1.27\, M$. 

\begin{figure}
\includegraphics[width=1\columnwidth]{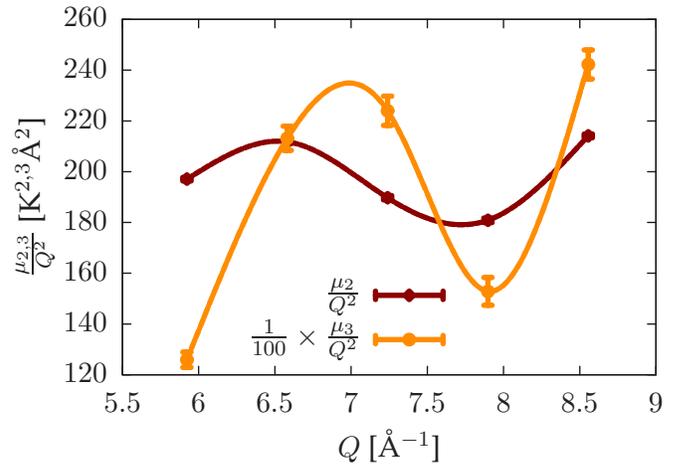} 

\caption{\label{fig:mu2_3y028} (Color online) Coherent oscillations of the
cumulants $\mu_{2}$ and $\mu_{3}$ appearing in the additive corrections
to the Impulse Approximation. Cumulants calculated at discrete wave-vectors
are interpolated by cubic splines as a guide to the eye.}

\end{figure}

In order to characterize the density excitations of $^{4}\mathrm{He}$
in this region of momenta, we have analyzed the dynamic structure
factor in terms of a cumulant expansion, Eq. (\ref{eq:cumulomega}),
thus extending the scope of previous theoretical results, \cite{Glyde:1985lr}
which were essentially unable to predict the response of the solid
in this intermediate regime. In Figure \ref{fig:yLhfit1028} the calculated
dynamic structure factor at the melting density $\rho_{1}$ is shown
along with its decomposition in terms of the leading Impulse Approximation
and its additive corrections of Eq. \eqref{eq:ExplCumul}. The coefficients
$\mu_{n}$ are taken as free parameters in the fitting procedure of
the dynamic structure factor, much as it is done in the analysis of
the experimental data.\cite{Glyde:1994fk} 

The bare Impulse Approximation of the dynamic structure factor overlooks
quantum coherence effects in the density response functions, which
are experimentally observed\cite{Martel:1976db,Tanatar:1987rw} in
superfluid $^{4}\mathrm{He}$. A better account of quantum coherence
can be achieved including higher order corrections in the cumulant
expansion of Eq. \eqref{eq:cumulomega} whose non-vanishing contribution
is in fact recognized in the fitting of our reconstructed spectra,
Figure \ref{fig:yLhfit1028}. The presence of quantum coherence between
Helium atoms in the solid is further appreciated upon looking for
deviations from the incoherent approximation of Eq. \eqref{eq:Sinco}.
A cumulant expansion of the incoherent dynamic structure factor can
be carried out, and it is known \cite{GlydeBook} that the cumulants
of such an expansion $\mu_{2}^{\textit{inc}}$ and $\mu_{3}^{\textit{inc}}$
increase monotonically as $Q^{2}$ whereas $\mu_{4}^{\textit{inc}}$
and $\mu_{5}^{\textit{inc}}$ increase monotonically as $Q^{4}$.
Characteristic $Q-$dependent oscillations in the ratios $\mu_{2,3}/Q^{2}$
and $\mu_{4,5}/Q^{4}$ can be therefore exploited to infer deviations
from the purely incoherent response. Such oscillations have been observed
in the superfluid phase\cite{Martel:1976db} and theoretically justified
within a T-matrix approximation of the He-He atom scattering.\cite{Tanatar:1987rw}
A similar behavior has also been recently observed in solid $^{4}\mathrm{He}$,\cite{Diallo:2004lr}
although no satisfactory \textit{ab-initio} theoretical description
exists yet. The cumulant dissection of the spectral properties extracted
from our QMC simulations is a natural tool to examine the relics of
quantum coherence in the intermediate wave-vector region, which closely
parallels the experimental analysis. In Figure \ref{fig:mu2_3y028}
we show the $Q-$dependent oscillations in the ratios $\mu_{2}/Q^{2}$
and $\mu_{3}/Q^{2}$ as found in our analysis of the dynamic structure
factor, which are a quite clear manifestation of the residual coherence
in the dynamics of solid helium in this intermediate region. The quantitative
aspects of this analysis may be influenced by the quality of the Maximum
Entropy reconstruction of the spectrum. However, the shift of the
peak position with respect to the free particle recoil frequency should
be reliable information, as suggested by the good agreement of the
calculated and measured phonon dispersions previously shown.

\section{\label{sec:Conclusions}Concluding remarks}

In this work we have demonstrated and successfully applied a complete
scheme to study the lattice dynamics of crystalline $^{4}$He at zero
temperature. Although the stochastic nature of QMC methods limits
us to explore the quantum imaginary time dynamics, we have shown that
quantitative accuracy can be nonetheless achieved. One of the most
appealing features of our analysis is the possibility to directly
parallel the experimental investigation based on neutron scattering.
The study of the full dynamic structure factor has allowed us to describe
both the phonon and the intermediate wave-length regions of excitations. 

At lower density, where the quantum fluctuations substantially affect
the dynamics, we have obtained satisfactory results for the phonon
energies, whereas approximate quasi-harmonic theories have shown difficulties
in the accurate determination of the vibrational dispersions. An interesting
point deserving more research is the effect of the adopted pair-potential
on the calculated phonon branches. Our study suggests a degradation
of the agreement to experimental data at higher density which could
probably be alleviated upon the inclusion of higher than two body
terms in the interatomic potential. 

In the intermediate regime of wave-vectors, where the density excitations
are better understood in terms of corrections to the small wave-length
behaviour, we have shown that residual coherence in the quasi-particle
excitations is present as in superfluid helium. Altough much of the
theoretical and experimental efforts on the merging between the phonon
and the single particle regimes have concentrated on the archetypal
quantum solid --$^{4}$He-- we believe that an extension of such an
analysis to other quantum solids such as $^{3}$He or molecular hydrogens
H$_{2}$ would surely be worthwhile. 
\begin{acknowledgments}
We gratefully acknowledge the allocation of computer resources at
the CINECA supercomputing center from the CNR-INFM \textit{Iniziativa
Calcolo per la Fisica della Materia} and financial support from MIUR
through the \emph{PRIN 2007} program. 

\end{acknowledgments}

\end{document}